# Ultrafast Measurements of the Interfacial Spin Seebeck Effect in Au and Rare-Earth Iron Garnet Bilayers

Víctor H. Ortiz[1,2,*], Michael J. Gomez[1,*], Yawen Liu[2], Mohammed Aldosary[2],

Jing Shi[2], Richard B. Wilson[1]

1) Department of Mechanical Engineering and Materials Science and Engineering Program, University of California, Riverside, California 92521, USA

2) Department of Physics and Astronomy Program, University of California, Riverside, California 92521, USA

*Denotes equal contribution.

We investigate picosecond spin-currents across Au/iron-garnet interfaces in response to ultrafast laser heating of the electrons in the Au film. In the picoseconds after optical heating, interfacial spin currents occur due to an interfacial temperature difference between electrons in the metal and magnons in the insulator. We report measurements of this interfacial longitudinal spin Seebeck effect between Au and rare-earth iron-garnet insulators, i.e. $RE_3Fe_5O_{12}$, where RE is Y, Eu, Tb, Tm. By systematically varying the rare-earth element, we modify the total magnetic moment of the iron-garnet. We use time domain thermoreflectance (TDTR) measurements to characterize the thermal response of the bilayer to ultrafast optical heating. We use time-resolved magneto-optic Kerr effect (TR-MOKE) measurements of the Au layer to measure the time-evolution of spin accumulation in the Au film. Replacing Y with other rare-earths enhances the electron-magnon conductance $G_{e\text{-}m}$ at the Au iron-garnet interface by as much as a factor of three. The electron-magnon conductance does not follow the trend of either the total magnetization of the iron garnet or the magnetic moment of the rare-earth.

# I. Introduction

The longitudinal spin Seebeck effect (LSSE) describes the injection of spin-currents into a nonmagnetic metal in response to a temperature gradient across a nonmagnetic-metal/magnetic-insulator heterostructure [1–3]. Spin-current across the metal/insulator interface occurs as a result of electrons in the metal flipping spin by emitting or absorbing magnons in the insulator. A temperature difference across the interface between electrons and magnons creates an imbalance in the number of emitted vs. absorbed magnons. The resulting heat-current is

$$j_Q = G_{em} \Delta T_{em}$$ (1)

where $G_{em}$ is the electron-magnon interfacial thermal conductance and $\Delta T_{em}$ is the temperature drop across the interface between electrons in the metal and magnons in the insulator. The heat-current is accompanied by a spin-current equal to the product of the heat-current and the ratio of angular momentum per unit of thermal energy [4]. The spin-current, in units of $A/m^2$, is

$$j_s = j_Q \frac{2e}{k_B T}.$$ (2)

The electron-magnon interfacial thermal conductance $G_{em}$ is a transport coefficient of central importance for spin Seebeck phenomena. The magnitude of $G_{em}$ is determined by the strength of quasi-particle interactions between electrons in the metal and magnons in the magnetic insulator. Despite its importance, there are few quantitative measurements of $G_{em}$.

Most experimental studies of the LSSE measure a transverse electrical current generated by the inverse spin Hall effect and spin-injection across the metal/insulator interface due to temperature gradients [5–11]. In addition to depending on $G_{em}$, these inverse spin Hall signals are sensitive to the energy-transfer coefficient between electrons and phonons in the metal layer [12], bulk spin-currents in the magnetic insulator [11] [12], the spin-diffusion length of the metal, and the inverse spin Hall angle of the metal [13–16]. The complex interplay of these phenomena makes it difficult to quantify $G_{em}$ from inverse spin Hall measurements. As a result, there is currently a lack of experimental data on how $G_{em}$ depends on the magnetic properties of the insulator. Another challenge to developing a fundamental understanding of $G_{em}$ is systematic differences in the design of experiments prevent quantitative comparisons of results from different research groups [16].

Ultrafast optical pump/probe techniques can provide a more direct measure of $G_{em}$ than steady-state inverse spin Hall effect experiments. Pump/probe experiments measure the spin Seebeck dynamics on femto- to pico-second timescales after ultrafast heating of the metal with a laser pulse [17,18]. On such short timescales, heat has not yet diffused from the metal into the magnetic insulator, i.e., $\nabla T$ in the insulator is approximately zero. As a result, ultrafast spin Seebeck effect experiments are not sensitive to magnon transport in the magnetic insulator [17,18]. The goal of our study is to examine how ultrafast spin Seebeck signals depend on the properties of the magnetic insulator.

We characterize the interfacial LSSE in Au/rare-earth iron-garnet bilayers. The rare-earth in our rare-earth iron-garnet (REIG = $RE_3Fe_5O_{12}$) layer is one of the following: Y, Eu, Tb, Tm. In the REIGs, the magnetic moment of the rare-earth is antiparallel to the three Fe ions sitting in the tetrahedral sites and parallel to the two Fe ions situated in the octahedral sites of the garnet. Consequently, interchanging the rare-earth ion from Y to Tm to Eu to Tb systematically decreases the total magnetic moment of the iron-garnet layer at room temperature [19,20].

To measure $G_{em}$ for these Au/iron-garnet bilayers, we combine time-resolved magneto optic Kerr effect experiments (TR-MOKE) [17] [21] and time-domain thermoreflectance experiments (TDTR) [22]. Both experiments use time-domain measurements of optical properties to determine the temperature and magnetic response of the sample to ultrafast optical heating of the Au layer. TDTR is a well-established method for measuring thermal transport properties and characterizing temperature fields that result from laser heating [22–24]. TR-MOKE measurements allow us to directly measure spin accumulation in the non-magnetic layer that results from interfacial spin-currents between the Au and iron-garnet layer (Eq. 1 and 2) [17,21].

## II.     Experimental methods

We grew REIG thin films by pulsed laser deposition (PLD) from densified ceramic targets. The target preparation methods are described in Ref. [25]. High quality, ultra-flat YIG, TbIG, TmIG and EuIG films, with thickness t ≈ 20 nm were deposited on (111)-oriented GGG (YIG and TbIG), (111)-oriented NGG (TmIG) and (001)-oriented GGG (EuIG) single crystal substrates. After ultrasonic cleaning in acetone, followed by rinsing in alcohol, the substrates were baked at 220°C for 5 hours under high vacuum (<$10^{-6}$ Torr) to reduce physically adsorbed molecules on the substrate prior deposition. Then, the substrates were annealed at 600°C in a 1.5 mTorr oxygen with 12% (wt. %) ozone atmosphere for 30 minutes. Under these conditions, a 248 nm KrF excimer laser pulse was set to strike the REIG target with a power of 135 mJ and a repetition rate of 1 Hz. After deposition, the samples were annealed *ex-situ* at 800°C for 300s under a constant flow of oxygen using a rapid thermal annealing (RTA) process, this process is required to obtain the single crystal structure in our iron garnet films [26]. The samples were then immediately loaded into a high vacuum magnetron sputtering chamber (AJA Orion) and, in an attempt to remove physical absorbed molecules, annealed at 200° C for 1 hour in in a mixture of 1 mTorr ultra-high purity oxygen and 10 mTorr argon. After the samples cooled to room temperature following the anneal, a ~60 nm Au layer was sputtered from a 1" target in a 3.5 mTorr Ar atmosphere with a power of 10 W.

In addition to the REIG samples described above that we prepared for spin-Seebeck effect measurements, we prepared other samples for materials characterization. These additional samples had thicknesses of either ~20 or 200 nm, and were grown with nominally identical conditions. On these duplicate samples, we performed reflection high energy electron diffraction (RHEED) of the REIG films (Figure 1a) after rapid thermal annealing to confirm the single crystal character of the iron-garnet thin films. We also performed atomic force microscopy (AFM). The films are atomically flat with low root-mean-square (RMS) roughness (<2 Å RMS for 20 nm films) and with no superficial defects (Figure 1b). For the 200 nm samples, the AFM measurements showed a slight increase in the roughness (<3 Å RMS). We performed X-ray diffraction (XRD) on 200 nm samples using a PANalytical Empyrean diffractometer with Cu $K_\alpha$ radiation and a Ni filter, at room temperature in 0.002° steps in the 2θ range of 10°-90° (Figure 1c). For the EuIG samples, two main peaks for EuIG and GGG were observed, corresponding to the (004) and (008) Bragg peaks. For the YIG, TbIG and TmIG samples, one main peak for REIG and one for the substrate (GGG for YIG and TbIG, NGG for TmIG) are observed, corresponding to the (444) Bragg peak, therefore confirming epitaxy and the single crystal structure. We observe no evidence of secondary phases in the $\theta - 2\theta$ scan. After RHEED, XRD, and AFM measurements, the 200 nm thick samples were coated with Au films for thermal property measurements.

Magnetic hysteresis curves ($M$ vs. $H$) of the iron-garnet films were collected with a vibrating sample magnetometer (VSM) at room temperature. Measurements were done with the applied magnetic field both perpendicular and parallel to the plane. The paramagnetic background from the substrates was subtracted from the raw data (Figure 1d). For the EuIG, TbIG and TmIG samples, a clear easy-axis / hard-axis loop can be observed for fields perpendicular / parallel to the plane, respectively. Strong perpendicular magnetic anisotropy in EuIG, TbIG and TmIG samples is caused by the magnetoelastic effect and lattice mismatch with the substrate [27]. For YIG, the easy-axis is in the in-plane direction because of a lack of strain on the GGG substrate. An out-of-plane magnetic field strength of $\sim 160 \; kA/m$ saturates the moment of the YIG thin-films in the out-of-plane direction.

For the TDTR and TRMOKE measurements, we used a pump-probe system built around a Ti:sapphire laser with a repetition rate of 80 MHz and pulse width of 700 femtoseconds [28]. An electro-optic modulator (EOM) modulates the pump laser at a frequency of $f_{mod} = 10.7$ MHz. A delay stage varies the arrival time of the pump laser pulses to the sample relative to the probe pulses. For TDTR measurements, the reflected probe laser is focused on a single photodiode to monitor pump-induced changes in reflectance. The photodiode is connected to an RF-lock in. We analyze the ratio of the in-phase and out-of-phase signal, $V_{in}/V_{out}$, measured by the RF-lock in with a thermal model [22]. The in-phase signal describes the temperature evolution on picosecond to nanosecond time-scales. The out-of-phase signal arises from pulse accumulation and describes the temperature response of the sample on times-scales of $1/f_{mod}$. For TR-MOKE measurements, the probe beam reflected from the sample is split into orthogonal polarizations and focused onto one of two photodiodes in a balanced photodetector. Changes in the polarization of reflected light cause an imbalance of power on the two photodetectors. An RF lock-in measures the difference in power on the two photodiodes at 10.7 MHz. Further details of our TDTR and TR-MOKE setup are in Ref. [28].

## III. Results

We performed TDTR experiments on the thick REIG films to characterize their thermal properties. TDTR measurements of the Au coated ~200 nm thick iron-garnet films provide a measure of the REIG thermal conductivity and Au/REIG interface conductance. In Figure 2, we show TDTR data along with thermal model fits for the 60 nm Au/200 nm TmIG/NGG sample. For this sample, we observed a thermal interface conductance between Au and the REIG of ~ $90 \frac{MW}{m^2 K}$ and a thermal conductivity of $\Lambda \approx 1.65 \frac{W}{m \cdot K}$. The thermal conductivity values of the other thick REIGs we measured were similar, see Table I. These values are lower than typical for these materials. For example, single crystal YIG has $\Lambda = 7.4 \frac{W}{m \cdot K}$ [29] (for comparison, at room temperature the thermal conductivity of amorphous $SiO_2$ is $1.6 \frac{W}{m \cdot K}$). The low thermal conductivity suggests significant crystalline disorder [30] [31], possibly due to non-stoichiometry. Crystalline defects reduce the thermal conductivity of insulators by decreasing phonon mean-free-paths [32]. Prior studies have reported non stoichiometric compositions for PLD grown REIG thin-films [33] [34].

We now discuss TDTR measurements of the thin-film iron-garnet samples with thicknesses between 20 and 40 nm. These are the samples that we also performed TR-MOKE measurements on. Since the REIG thicknesses are small, TDTR data is not independently sensitive to the film thermal conductivity vs. the interface thermal resistance. Instead, TDTR measures the total thermal resistance between the Au and the gallium-garnet substrate. This total resistance includes three thermal resistances that add in series: the Au/iron-garnet interface resistance, the thermal resistance of the iron-garnet film, and the iron-garnet/gallium-garnet interface. We found a total thermal resistance (in $10^{-8} \frac{m^2 K}{W}$) between the Au and substrate of ~2.5 (TmIG), 2.3 (TbIG), 3.1 (EuIG) and 2.3 (YIG). From our TDTR experiments on ~200 nm thick iron-garnet films, we know the Au/REIG conductance and REIG thermal conductivity for TmIG, TbIG, EuIG and YIG. Assuming these values are unchanged for 20 nm thick samples, we estimate the interface conductance between the TmIG, TbIG, and YIG films and the substrate to be $\approx 300 \frac{MW}{m^2 K}$. The uncertainty for this value is significant, because the thermal resistance from the iron-garnet/substrate interface is small compared to the other two thermal resistances.

We report the results of our TR-MOKE measurements of the interfacial spin Seebeck effect in Figure 3. Optical heating of the Au film with a ~700 fs pump pulses causes a transient magnetic moment in the Au that persists for ~ 5 ps. We used polar MOKE to detect the out-of-plane magnetic moment of the Au film with a probe beam energy of 1.58 eV. The optical penetration depth of Au at 1.58 eV is ~13 nm [35], much less than the 60 nm Au film thickness, while reported values for the spin diffusion length of Au are between 30-100 nm [36]. Therefore, the TR-MOKE measurements are only sensitive to the magnetic moment of the Au film, and not the iron-garnet film underneath. We control the orientation of the REIG moment by applying a 240 $kA/m$ out-of-plane external magnetic field. Our experiment measures only the out-of-plane magnetic moment. However, we expect the magnetic moment of the Au layer to be zero in other directions. The source of the spin accumulation in the Au is demagnetization of the iron-garnet, whose moment points in the out-of-plane direction at all times. The Kerr rotations reported in Figure 3 are deduced from the difference in signal measured with positive vs. negative magnetic field [28]. The data in Figure 3 was collected with an incident pump with spot size of 7.5 μm and fluence of 5 $J/m^2$ .

## IV.    Analysis

We observe that the TR-MOKE signal rises within ~500 fs of the pump laser excitation of the Au and persists for ~ 5 ps. To understand this behavior, we follow Ref. [17], and use a two-temperature model to describe the thermal response of the system to heating and a spin-diffusion equation to describe spin dynamics in the metal.   The heat equations for the electrons and phonons in the Au layer are

$$C_e \frac{\partial T_e}{\partial t} - \Lambda_e \frac{\partial^2 T_e}{\partial x^2} = g_{e-ph}(T_{ph} - T_e) + p(z,t),$$

$$C_{ph} \frac{\partial T_{ph}}{\partial t} - \Lambda_{ph} \frac{\partial^2 T_{ph}}{\partial x^2} = g_{e-ph}(T_e - T_{ph}).$$

(3)

Here, $T_e$ and $T_{ph}$ describe the temperature rise of electrons and phonons above ambient, $C$ is volumetric heat capacity,  $\Lambda$ is the thermal conductivity, $g_{e-ph}$ the energy transfer coefficient between electrons and phonons, and $p(z,t)$ the optical absorption profile determined by a transfer matrix optical model [17]. We use two coupled heat equations similar to Eq. (3) for the REIG, but that describe phonon and magnon thermal reservoirs instead of phonon and electron thermal reservoirs. Finally, we assume the thermal response of the substrate is described by a single heat-equation for phonons. We apply adiabatic boundary conditions on the Au electrons and phonons at the surface. We assume the electron thermal reservoir is coupled to the spin thermal reservoir by an electron-magnon conductance, see Eq. 1. We assume the phonon thermal reservoirs of adjacent layers are coupled by the interface conductance values derived from the TDTR measurements described above. We set the thermal conductivity of the Au electrons to $\sim 200\, \frac{W}{m \cdot K}$ based on four-point probe measurements of the electrical resistivity and the Wiedemann Franz law. Other parameters for Au are taken from Ref. [37]. We set the magnon heat capacity of the iron-garnet layer to $1.2 \times 10^4\, \frac{J}{m^3 K}$ [38], and we fix the magnon thermal conductivity to $1 \times 10^{-1}\, \frac{W}{m \cdot K}$ [12] and magnon-phonon coupling parameter to $3 \times 10^{14}\, \frac{W}{m^3 K}$ [17]. We assume these parameters are independent of the rare-earth composition of the sample. All thermal model parameters in Eq. (3) are fixed except $G_{em}$. We treat $G_{em}$ as a fit parameter and is used to fit model predictions and TR-MOKE data (discussed in more detail below). The predictions for the electron and phonon temperatures in the Au layer following laser irradiation are shown in Fig. 4a. During laser excitation, the Au electron temperature increases by $\sim 100K$, and then thermalizes with the lattice via phonon emission [39] with a relaxation time of 1.3 ps.

The large interfacial temperature difference between the Au electrons and the iron-garnet layer causes spin-injection into the Au layer (Eqs. 1-2). We describe the resulting spin accumulation in the Au layer with the spin diffusion equation

$$D \frac{\partial^2 \zeta_S}{\partial x^2} - \frac{\partial \zeta_S}{\partial t} - \frac{\zeta_S}{\tau_S} = 0 \tag{4}.$$

Here, the spin accumulation $\zeta_S$ is the difference is chemical potential for up and down spins, the diffusivity of the spin is $D = \Lambda_e / C_e$ and $\tau_s$ the spin relaxation time. The spin current at position $x$ is $j_s = \frac{\sigma}{2e} \frac{\partial \zeta_S}{\partial x}$, where $\sigma$ is the electrical conductivity. We apply $j_s = 0$ as a boundary condition at the Au film surface and use Eq. (2) as the boundary condition at the Au/iron-garnet interface. Together, Eqs. 1-4 allow us to predict the spin accumulation in the Au layer as a function of time. To compare with the TR-MOKE data, we assume a Kerr rotation of $24 \times 10^{-9} \frac{rad}{A/m}$, as reported in Ref. [17]. We assume the TR-MOKE experiment measures the spin accumulation at the surface of the Au layer. The parameters $\tau_S$ and $\alpha = \frac{2e}{k_B T} G_{e-m}$ are determined by fitting model predictions to the experimental data in Figure 3a. As the magnon thermal properties have significant uncertainty, to determine if this uncertainty affects our results, we report a sensitivity analysis [40] for the magnon thermal conductivity ($\Lambda_m$), magnon heat capacity ($C_m$) and magnon-phonon coupling ($g_{m-ph}$). For these magnon thermal properties, we find changes as large as 2 orders of magnitude have only a small effect on the best fit values of $G_{em}$ and $\tau_s$. The best-fit values for $G_{em}$ and $\tau_s$ are most sensitive to the thermal properties of the Au electrons, which are better known then the magnon thermal properties.

## V.    Discussion

We observe an increase of the magnetic moment into the Au layer due to spin injection caused by the thermal gradient (Figure 3b). Reflected probe light rotates in the clockwise (negative) direction when the applied external field points in the positive direction, i.e. when the field points away from the sample surface. Au has a negative Kerr angle at 783 nm. Therefore, the spin accumulation in Au is in the same direction to the applied field, which corresponds to the $Fe^{3+}$ in d-sites (tetrahedral) of the REIGs, and opposite to the $Fe^{3+}$ in the a-sites (octahedral) and the $RE^{3+}$ in the c-sites (dodecahedral). This allows us to conclude that the major contributor to the spin current corresponds to the Fe atoms in the tetrahedral sites of the iron garnet.

To confirm the relationship between signal sign and orientation of the moment in our experiments, we did several control measurements. First, we measured the static Kerr angle of a 130 nm Ni thin film at 783 nm. We observed a negative Kerr angle for Ni, i.e. clockwise rotation of the polarization for positive magnetic field, in agreement with prior studies [41]. We then performed the TRMOKE measurements of the Ni film with the same conditions that we used for our Au/REIG samples. Since pump heating decreases the magnetic moment and decreases Kerr rotation, transient MOKE signals of ferromagnetic metals have the opposite sign as static Kerr measurements. As expected, we observed positive time dependent MOKE signals from the Ni thin film. Finally, to confirm the sign of the Kerr angle of Au is negative, we performed inverse Faraday effect measurements on a 20 nm Au film on sapphire [42]. Upon photoexcitation with right circularly polarized pump light, we observed a positive Kerr rotation of the probe beam. Right circularly polarized pump light induces a magnetic moment in the Au in the negative direction (right-hand rule), so we conclude Au has a negative Kerr angle.

We qualitatively sketch the spin Seebeck effect dynamics predicted by Eqs. 1-4 in Figure 4. Before the pump laser strikes the sample, the Au/REIG bilayer is in thermal equilibrium and the Au layer has no magnetic moment (Figure 4bi). Prior to thermalization with the phonons, the thermal diffusivity of the electrons is large, $\sim$ 10 mm$^2$/s [43], due to the small heat-capacity of the electrons. So, heat diffuses rapidly across the film thickness on a time-scale of $C_e d_{Au}^2/\Lambda_e \approx 0.4$ ps. The increase in electron temperature at the Au/iron-garnet interface increases the rate of interfacial magnon emission and drives an interfacial spin-current. Spin accumulation causes the magnetic moment of the Au layer to increase (Figure 4bii). After ~2 ps, the interfacial spin-current decreases due to a lower electron temperature, and spin-flip scattering relaxes the spin accumulation in the Au layer (Figure 4biii).

The spin accumulation for the different Au/REIG bilayers are shown in Figure 3a, with the corresponding fitted parameters in Table I. The values we report (in $\frac{MW}{m^2 K}$) are 0.9 (YIG), 1.6 (EuIG), 2.1 (TbIG) and 3.1 (TmIG). The value obtained for our Au/YIG bilayer is consistent with prior observations [17].

It is notable and somewhat surprising that $G_{em}$ of the YIG samples is smallest despite YIG possessing the largest magnetic moment at room temperature. As mentioned before, replacing the Y$^{3+}$ ion in the dodecahedral sites with a rare-earth element with a significant magnetic moment decreases the total magnetic moment [19] [20]. At room temperature, the accepted values for our REIGs magnetization are (in $kA/m$) 141 for YIG, 110 for TmIG, 95 for EuIG, and 19 for TbIG [20,44]. The room-temperature moments are in part a consequence of the contribution for the rare-earth elements in 3+ state in our materials: $\mu_Y = 0$ $\mu_B$, $\mu_{Tm} = 7.56$ $\mu_B$, $\mu_{Eu} = 3.4$ $\mu_B$, and $\mu_{Tb} = 9.72$ $\mu_B$. It is clear that the obtained electron-magnon interfacial conductance values do not follow the trend of either the total magnetization or rare-earth magnetic moment. One possibility is that differences in $G_{em}$ are due to structural differences at the interface. However, our thermal transport measurements do not provide evidence for this hypothesis. The interface conductance is sensitive to interfacial bonding and interfacial disorder [45], and the phonon thermal interface conductances we derive from TDTR experiments are similar in all samples (Table I). Another possibility is that the differences in $G_{em}$ for the various REIGs are related to differences in the magnon dispersion. Geprägs *et al.* explain the temperature dependence of the spin Seebeck effect in gadolinium iron garnet insulator [7] by considering the details of the magnon spectrum. They concluded high energy exchange magnons play a key role. The frequency of high energy exchange magnons will be strongly affected by substituting yttrium with rare-earth elements with different magnetic moments. However, a detailed analysis of how rare-earth substitution effects the magnon spectrum is beyond the scope of the current experimental study.

To put the electron-magnon conductance values we observe in context, we compare to the electron-phonon energy transfer coefficient in Au. If we assume a volume of interaction near the interface of ~ 1 nm, an electron-magnon conductance per unit area of $3 \frac{MW}{m^2 K}$ is equivalent to a volumetric energy transfer coefficient of ~ $3 \times 10^{15} \frac{W}{m^3 K}$ between Au electrons and REIG magnons. This value is roughly ten times lower than the energy transfer coefficient between Au electrons and Au phonons. The electron-phonon energy transfer coefficient in Au is proportional to the rate of spontaneous phonon emission by a hot electron [46]. Therefore, we estimate the rate of REIG magnon emission by hot electrons in proximity to the interface is roughly one order of magnitude lower than the rate of Au phonon emission.

In summary, by performing a series of pump/probe measurements, we characterized the longitudinal spin Seebeck effect in Au/REIG bilayer systems on sub-picosecond timescales. Typically, the spin Seebeck effect is measured under steady-state conditions. By using a series of TDTR and TR-MOKE measurements, we characterize the ultrafast interfacial spin Seebeck effect in Au/REIG layers with very similar interface morphology. We observe a considerable enhancement in the electron-magnon conductance $G_{em}$ of EuIG (~1.5X), TbIG (~2X) and TmIG (~3X) samples compared to YIG samples. Our results imply the strength of interfacial interactions between Au electrons and iron-garnet magnons is approximately an order of magnitude weaker than interactions between Au electrons and Au phonons. Our findings are important for ongoing efforts to develop a fundamental understanding of interfacial electron-magnon interactions in magnetic heterostructures [47].

This work was primarily supported by the U.S. Army Research Laboratory and the U.S. Army Research Office under contract/grant number W911NF-18-1-0364. V.O., Y. L, M. A. and J. S. also acknowledge through SHINES, an Energy Frontier Research Center funded by the US Department of Energy, Office of Science, Basic Energy Sciences under Award No. SC0012670.

## Data availability

The data that support the findings of this study are available from the corresponding author upon reasonable request.

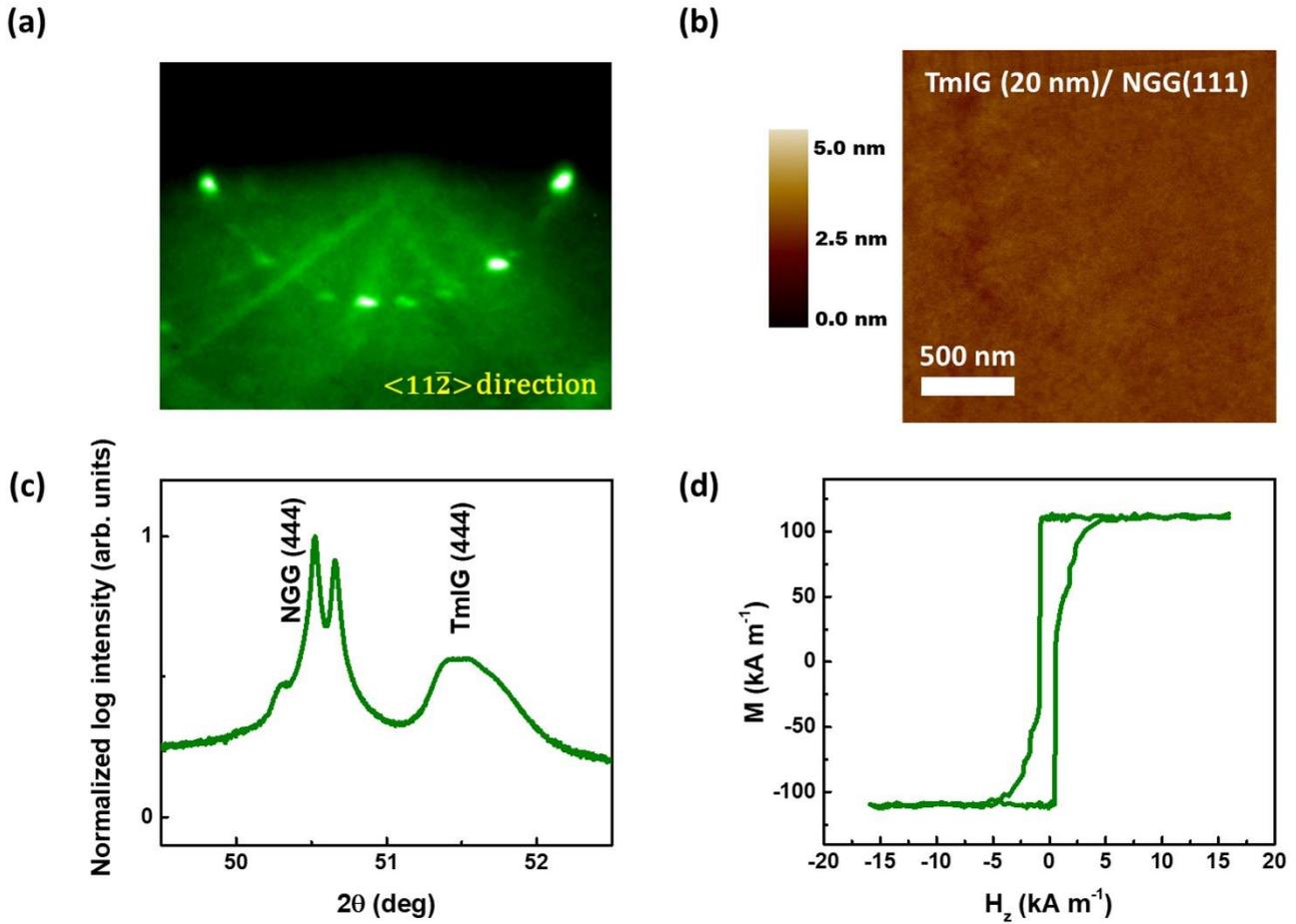

***Figure 1*** *Structural, morphological and magnetic characterization of TmIG films. a)RHEED pattern along the ⟨11$\bar{2}$⟩ direction for 20 nm TmIG/NGG(111) thin film after RTA treatment, showing single crystal structure. b) AFM image of the 20 nm TmIG/NGG(111) sample, the measured roughness is 1.53 Å RMS. c)XRD θ-2θ scan for the 200 nm TmIG/NGG(111) sample, clearly showing the corresponding peak for the (444) orientation for both substrate and film. d) M vs H hysteresis loop for field perpendicular to the film plane for the 20 nm TmIG/NGG(111) sample, showing a strong PMA behavior.*

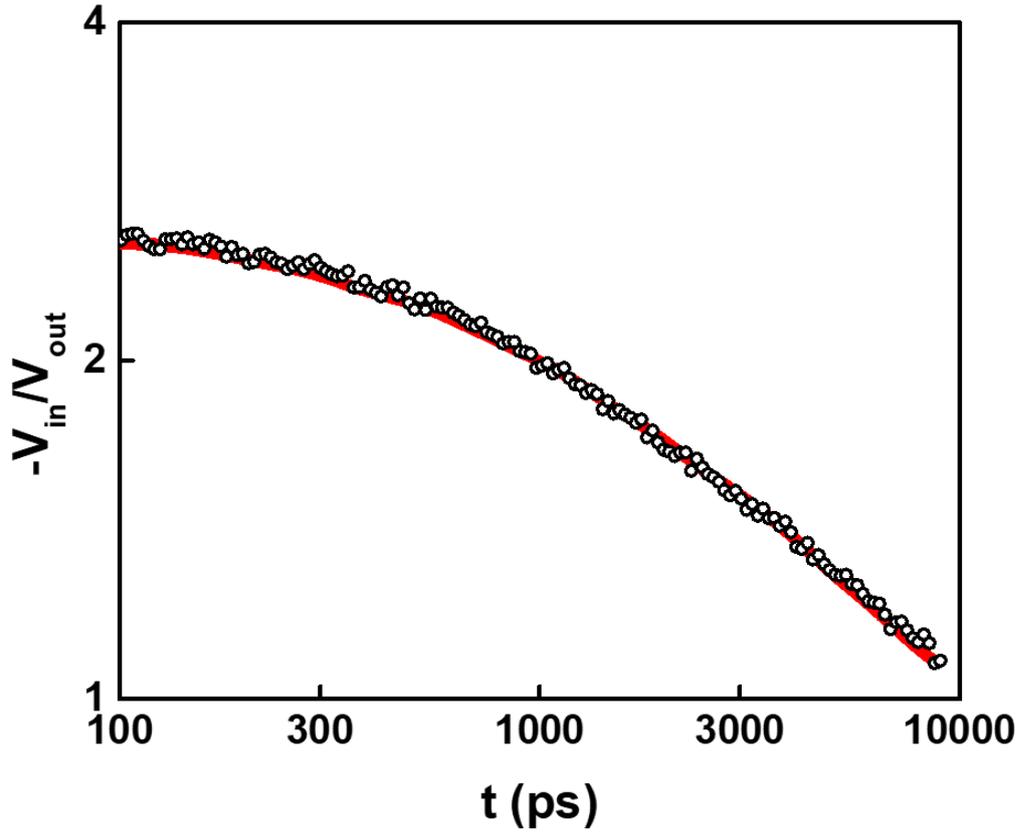

*Figure 2* Thermoreflectance data and model prediction for 60 nm Au/200 nm TmIG/NGG(111) sample. A TDTR measurement was performed on the sample, the experimental ratio of the in-phase and out-of-phase data were fit with a thermal model (red line) in order to obtain the interfacial thermal conductance $\left(95 \frac{MW}{m^2 K}\right)$.

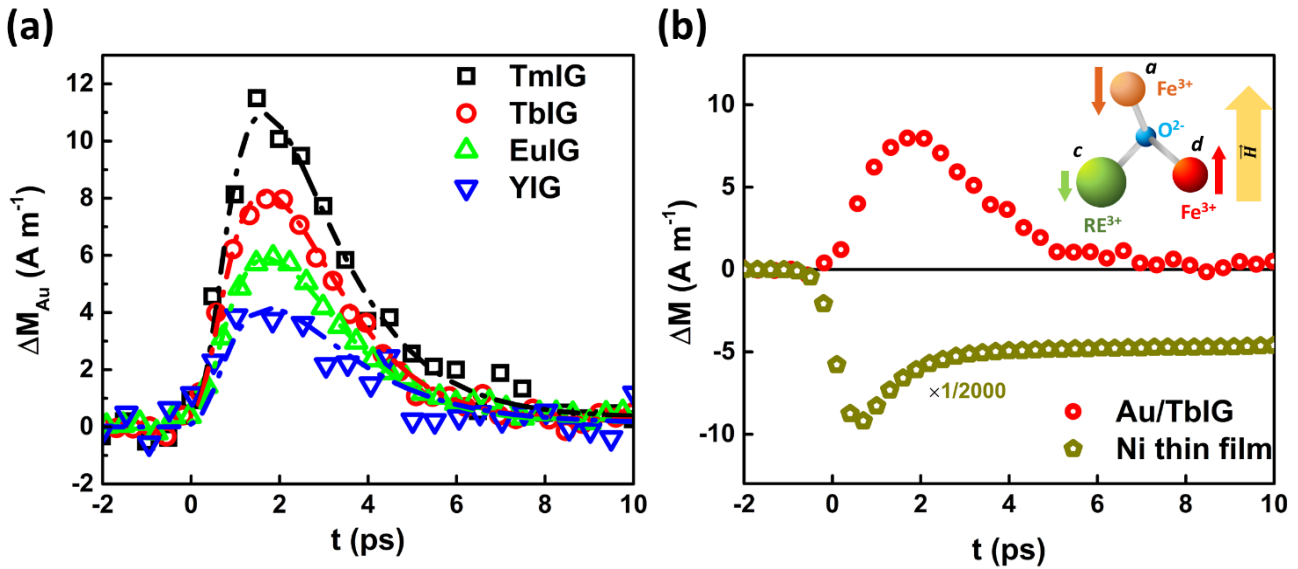

*Figure 3* a) TR-MOKE of 60nm Au as a function of iron garnet spin source. 60nm Au sputtered onto thulium (black), terbium (red), europium (green), and yttrium (blue). The amount of spin accumulation after femto-second laser absorption in the Au layer depends strongly on the type of iron garnet. The dashed lines correspond to the fitted curves from the spin diffusion and thermal models. b)Determination of the spin accumulation sign in the Au layer: a Ni thin film control sample presents a negative TR-MOKE signal due to temperature demagnetization, under the same measurement conditions the Au/TbIG sample displays a positive TR-MOKE signal due to the spin injection onto the Au layer. This proves that the spin accumulation in the Au layer has the same direction as the applied field, which corresponds to the $Fe^{3+}$ d-sites (tetrahedral).

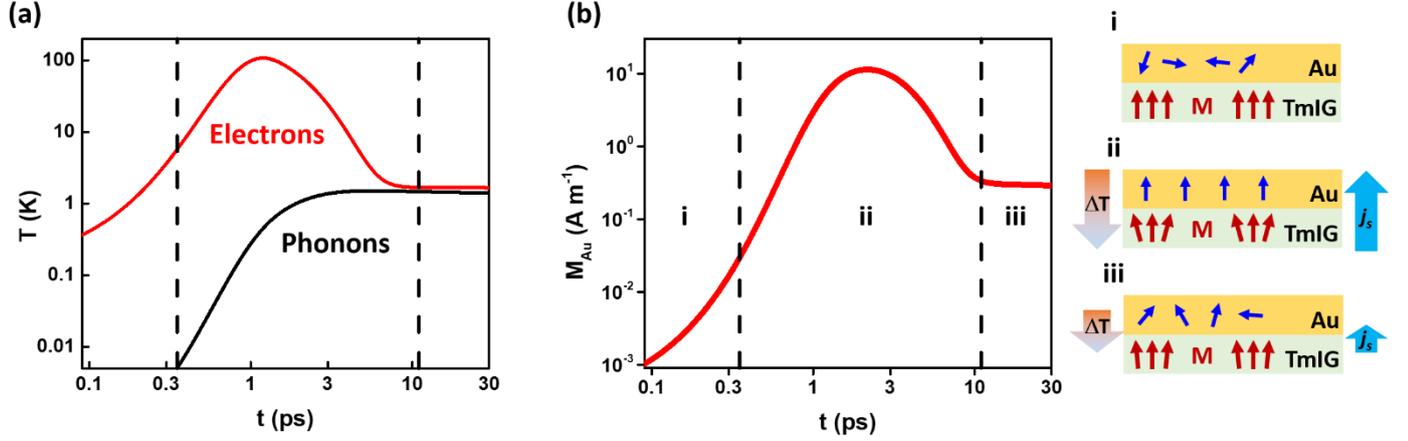

***Figure 4*** *a) Temperature rise of Au electrons and TmIG phonons, calculated using the two-temperature model. b)Time evolution of the spin accumulation in Au: i) Au and TmIG layers are in thermal equilibrium, Au shows no magnetic moment; ii) a laser-induced temperature gradient generates a spin accumulation due to SSE, causing an increase of magnetic moment in the Au layer; iii) spin accumulation in Au layer decays due to spin-flip scattering.*

| Sample | $\Lambda_{film}$ (MW m$^{-2}$ K$^{-1}$) | $\Lambda_{int\ Au/REIG}$ (MW m$^{-2}$ K$^{-1}$) | $\Lambda_{int\ REIG/substrate}$ (MW m$^{-2}$ K$^{-1}$) | $R_{sample}$ ($\times 10^{-5}$ W$^{-1}$ m$^2$ K) | $\alpha$ ($\times 10^8$ A m$^{-2}$ K$^{-1}$) | $\tau_s$ (ps) | $G_{e-m}$ (MW m$^{-2}$ K$^{-1}$) |
|---|---|---|---|---|---|---|---|
| YIG | 1.45 | 110 | 300 | 2.30 | 0.70 | 1.40 | 0.9 |
| EuIG | 1.80 | 90 | 100 | 3.13 | 1.20 | 1.50 | 1.6 |
| TbIG | 1.92 | 85 | 300 | 2.31 | 1.60 | 1.60 | 2.1 |
| TmIG | 1.65 | 95 | 300 | 2.45 | 2.40 | 1.35 | 3.1 |

***Table I*** *Thermal properties and spin-diffusion fitted parameters for Au/REIG thin films.*